# An Ant Based Framework for Preventing DDoS Attack in Wireless Sensor Networks

Dimple Juneja
Maharishi Markandeshwar University,
Mullana, Haryana, India
Email id: dimplejunejagupta@gmail.com

Neha Arora
Banasthali University, Banasthali,
Rajasthan, India.
Email id: aroraneha21@gmail.com

**Abstract**

Security and Privacy are two important parameters that need to be considered when dealing with Wireless Sensor Networks as WSN operate in an unattended environment and carry sensitive information critical to the application. However, applying security techniques that consume minimum resources is still a challenge and this paper makes an attempt to address the same. One of the major attacks in sensor network is Denial of Service(DoS) attack that not only diminishes the network capacity but also affects the reliability of information being transmitted. This work is an extension of our previous work which could successfully detect DDoS using ants. However, no emphasis was made towards the prevention mechanism. In this paper an ant-based framework that exploits the significance of stateless and stateful signatures and hence preserving the legitimate packets only, thereby discarding the contaminated packets has been proposed.

## 1. Introduction

Need of hour is sensor based computational structure that can gather information disseminated even in remote and inaccessible areas. Wireless Sensor Network (WSN) can be readily deployed and are well adapted to monitor the activities such as in military applications; a researcher can sit miles away and watch the activities of an active volcano, habitat monitoring, health monitoring, etc. with an additional advantage of being inexpensive. But due to the resource constraint nature of sensor networks, these lack security which is critical in many applications such as military sensing and tracking. It is this resource constraint nature that makes a variety of DDoS attacks easy in sensor network. Traditional security mechanisms such as public key cryptography, authentication, etc involves lot of computation time and delays; thus result in decreased network performance and consumption of nodes energy because of which they are not very well suitable for application in sensor networks.

In this paper we propose a framework that can be deployed in already existing networks and can prevent the faked messages from being spread across the entire network. Pre-authentication filters are applied before actual verification of bogus messages.

The contributions of this paper are: Firstly, it doesn't require extra computational load as only small portion of packets compared with total throughput of network are analyzed. Secondly, without wasting much energy it saves the network from attacks, thus resulting in increasing the lifetime of network and trustable communication. Thirdly, the problem of false alerts is addressed neglecting which may itself lead to DoS attack.

The rest of the paper is organized as follows. In section II, related work is discussed. Section III describes the proposed system in detail. Finally Section IV discusses conclusions and future work.

## 2. Related Work

This section presents the work of eminent researchers in the field, highlighting the challenges in the existing solutions.

Flooding DoS attack poses a great threat as it generates large volume of traffic that prevents the legitimate user from accessing the service. It causes the links to be blocked and nodes to crash resulting in decreased network performance and even more sensors become useless due to depletion of energy in sending useless packets. A number of approaches have been proposed to counter the attack.

Wood *et. al.* in [1] has summarized different DoS attack and their effect on the sensor network. They have listed various possible attacks in each layer which tells the importance of security features in sensor nodes. Studies [2,3] shows the survivability of wireless adhoc networks in term of link connectivity and stability between sensor nodes but they lack in considering the security of sensor network. For instance, Wang [4] and Ali et.al. [5] proposed secure packet transfer using encryption, decryption and authentication of packet header but the performance of PKC is not yet good due to resource constraint nature of sensor network. Chiang *et al.* [13] proposed architecture by adding duplicated hardware by which the reliability and availability of sensor networks can be increased but redundant hardware requires additional costs.

Researchers [6, 7] proposed security mechanisms against DoS attacks but the proposed solutions cannot handle wide range of DoS attacks.

C. Meadow [21] proposed stronger authentication between communicating parties across a network but it while attempting to prevent DoS leaves itself open to attack due to high computational load required to defend the attack wherein [22] uses payment approach and  assumes that node willing to access resources would have to pay charges according to the level of service needed. This approach does not provide total solution as legitimate nodes refusing to pay would be denied of accessing service.

Authors in [8, 9, 10] use congestion algorithms to detect upsurges in traffic that can give rise to DoS but these approach may apply only simplistic signatures and also requires state information to be held on the nodes which is not a feasible solution in sensors because of limited memory. Shyne and Sterne in [11, 12] uses statistical monitoring to detect upsurges in traffic of a particular type and raise alert if something unusual is detected. Here a single alert can notify about many attack packets but it requires human intervention to monitor upsurges so is inefficient.

A critical look at the literature highlights the fact that although lot of work has been done towards the security of WSN however, nothing has proved to be so significant so as to be considered as best. Moreover, researchers have ignored the fact that software agents especially ants can be used as security staffers and can provide a protection against DDoS in WSN.  The upcoming section aims to propose an ant-based framework that would be able to achieve already stated objectives.

## 3.  Proposed Work

This section presents the high level view and working algorithm of the proposed architecture that considers a heterogeneous system with some nodes having more processing and battery power than other nodes. These nodes are referred to as adjunct nodes which are used to monitor the state of network and take appropriate actions when required. The adoption of heterogeneous system ensures a cost effective design and more effective implementation of overall application. The framework basically comprises of DDoS detecting Ants (DDA), Traffic Monitoring Ants (TFA) and Filtering Module where DDA is responsible for detecting DDoS attack and TFA applies stateful and stateless signature. Stateful signature analysis uses the algorithm proposed in [15] where reliability > 1 implies Flooding attack. For the confirmation of attack, the destination address of sample packets are compared, if they match it is likely that large volume of traffic is heading towards same destination which provides stateful signature of DoS attack. On the other hand stateless signature analysis performs pattern matching. It looks for particular string in the packet that is an indication of attack. The Filtering Module responds to the packets inspected. It tells whether an attack detected is just a false positive or truly an attack and the appropriate action to be taken to prevent it.

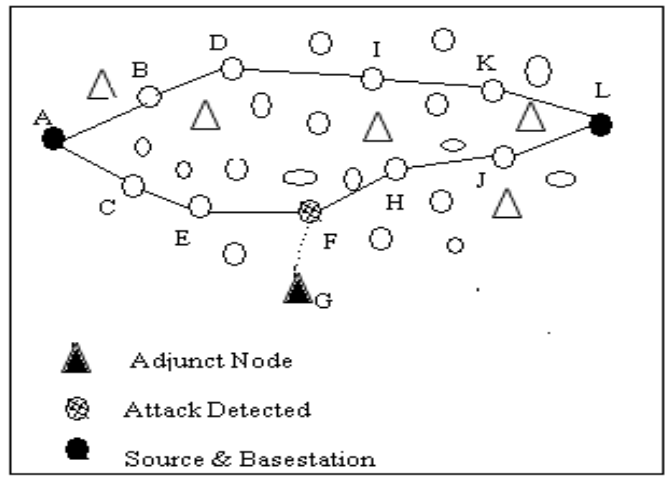

Figure 1: High Level View of Proposed Work

Figure 1 depicts the high level view of the prevention process during DDoS attack. The explanation is as follows:

i) Node E launches a flooding attack as part of distributed DoS attack against node F.
ii) The ant agent at node F notices rise in network traffic using algorithm proposed in [15] so it issues an alert message to monitoring agent sitting at the adjunct node G.
iii) The traffic is then directed to node G for analysis for the confirmation of attack using filtering module located at node G.
iv) As the attack gets confirmed the packets are dropped and an alert message containing signature of detected attack is sent back so as to drop packets on previous nodes as well.
v) This mechanism helps even in tracing the source of attack which can then be removed from the network so that it can't cause any further harm.
vi) This prevention process reduces computational overhead as follows: Firstly, only small portion of packets that exceed threshold value are inspected as compared to analyzing every packet. Secondly, the packets that are not a part of attack are redirected on appropriate route thus saving bandwidth that would have been wasted if they were dropped. Thirdly, the already existing detection mechanism is used to prevent the attack. As in the detection mechanism, congestion and attack were reported these acts as signature which are further analyzed by filtering module.

Following subsections illustrates the working of each module.

*3.1 DDoS Detecting Ants (DDA)*

They gather information about state of network. The ants have unique properties such as robustness, distributed problem solving and decentralized approach. It can very well adapt itself to network changes and can communicate with other ants to help solve complex problem through collaborative effort. As they can learn the network and environment if something unusual happens it can notice and can easily take appropriate action. The ants here carry data packets and check the status of each node on the way to destination. If in case it notices that the node it is visiting is unreliable i.e. congestion, link failure, flooding or any other form of DoS attack, it issues an alert signal to nearby TFA to take appropriate action. For details about DDA, refer [15].

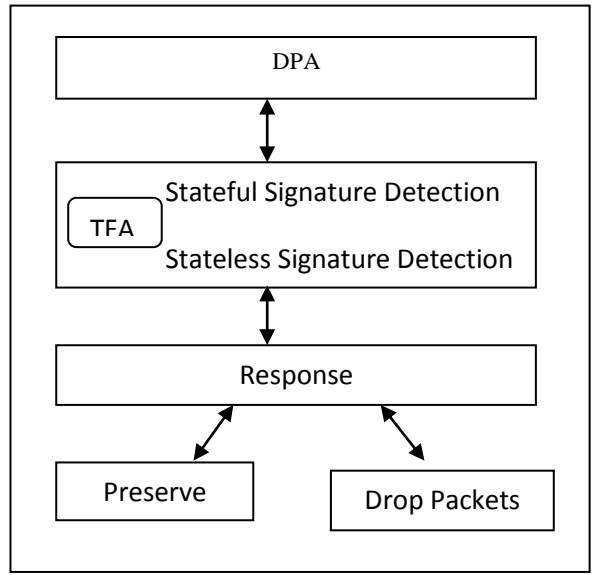

Figure 2: Filtering Module

### 3.2  Filtering Module

The idea behind applying filters is that DDoS attack directs a large volume of malicious traffic on links which can be easily filtered by pattern matching scheme without requiring much processing power to monitor each and every packet closely. Thus it would result in minimum delays and maximum throughput of network if only the suspicious traffic is analyzed.

The filtering module as shown in Figure 2 is situated on an adjunct node and comprises of 3 sub modules, namely: DDoS Preventing Ant (DPA), Traffic Filtering Ant (TFA) and Response.  The functionalities of each module are listed in Table 1.

Table 1: Functionalities DPA, TFA and Response Module

| Module | Responsibility |
|---|---|
| DDoS Preventing Ant (DPA) | • Responsible for communication between the sensor node and adjunct node.<br>• Upon high traffic DPA activates TFA at adjunct node & passes sample packets for analysis.<br>• Interacts with response module to know the result of analysis & send a report to Sensor node. |
| Traffic Filtering Ant (TFA) | • Responsible for confirming the attack.<br>• Stateful and stateless signatures are applied in parallel to the sample packets.<br>• Stateful signature alerts about high traffic heading towards a node<br>• Stateless signature verifies that an attack is actually taking place.<br>• The packets that are detected by both i.e common packets are dropped and rest are sent back as they are not a part of attack. |
| Response | • Takes the appropriate action i.e. whether to drop or preserve packets.<br>• In case of attack, it starts dropping packets unless the amount of incoming traffic reaches below threshold.<br>• Interacts with the DPA to send alert message to sensor node to drop packets on node itself.<br>• Also it tells the appropriate time when to lift dropping. |

**Flowchart and Working Algorithms**

The flowchart illustrating the proposed work is given in Figure 3. This procedure is activated only after the DDoS has been detected [15] at a particular node so that the attack doesn't prohibit the access of entire network and render the network useful for nothing. The major significance of this approach is that nodes can carry on with normal proceeding and there is no need of wasting time and energy in changing the route. Attacked traffic is diverted on a heterogeneous node which confirms the alert is truly an attack or just a false alert.

Whenever, an attack is detected the DDA residing on a node raises the alert i.e. whenever nodes reliability change its state from 1 to something else, an alert is raised. Also when buffer size of a node exceeds threshold limit, DDA detects that there is an unusual rise in network traffic. One possible reason of this rise can be flooding so in order to get confirmed it sends a sample of packets for analysis and also meanwhile starts searching for an alternative route that can be opted if due to attack the current route gets choked. These ants communicate with another set of ants available at the adjunct node referred to as DPA. Upon the activation of DPA, sample of packet is sent for analysis. DPA notices the difference between the current numbers of arriving packets on the node with sample packets whose number was greater than the threshold value. There can be three possible cases:

- Case 1: If both are equal that means their can be a possibility of attack so it waits a random amount of time and again compares the value.
- Case 2: If the current number is less than the sample number, then DPA claims that there is no attack and so packets are sent back to DDA without analysis and normal proceeding are carried out.
- Case 3: In reverse of case 2, DPA calls the filtering module and packets are analyzed using stateful and stateless signature in parallel. The common packets are dropped and an alert containing signature of attack is sent to DPA which in turn communicates with DDA. DDA on current node informs the previous node on the route from whom packets were received and asks it to drop the packet there only. This process goes on until the source from whom attack packets were generated is reached.

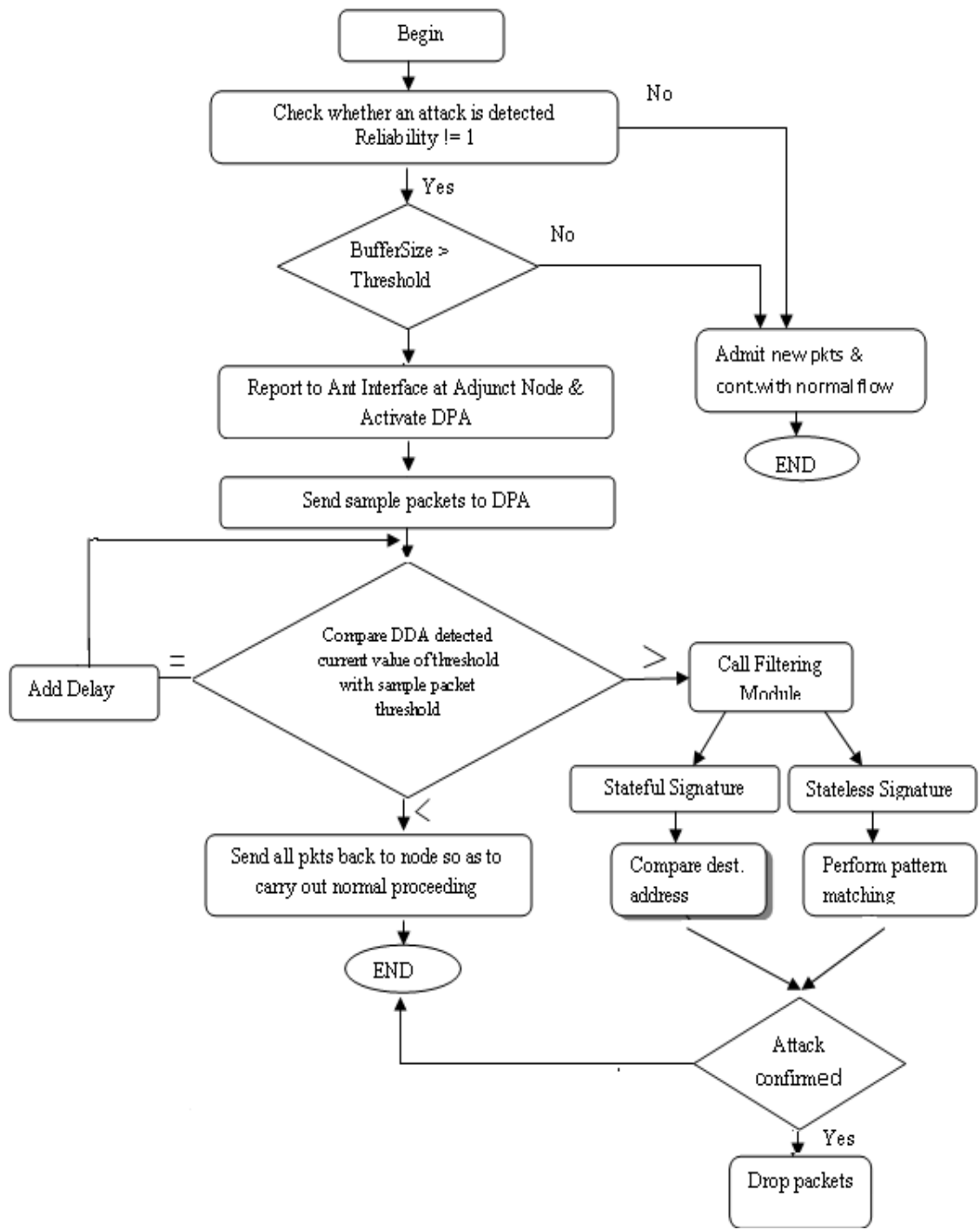

Figure 3: Flowchart of Proposed Work

Thus the mechanism helps in preventing and tracing the source of attack as well.

Pseudocode for the procedure illustrated above:
1. *while(reliability != 1)*
2. *{*
3. *dowhile(buffersize > threshold)*
4. *{*
5. *Activate DPA;*
6. *Send sample_packets to DPA;*
7. *Output=Compare current_threshold with sample_threshold();*
8. *}Endwhile*
9. *}Endwhile*
10. *End*

   7.1  *Compare current_threshold with sample_threshold()*
   7.2  *if(current_value of threshold < sample_packets threshold)*
   7.3  *output=Send packets back to node;*
   7.4  *Continue with normal proceedings;*
   7.5  *elseif(current_value of threshold > sample_packets threshold)*
   7.6  *Filtering module();*

```
7.7         elseif(current_value of threshold = sample_packets threshold)
7.8              output=Add delay;
7.9              goto step 7;

7.61    Filtering module()
7.62        do
7.63        {
7.64            if(dest_addr[i]==dest_addr[j])↕(pattern[i]==pattern[attack])//i,j€packet
7.65            Drop packets;
7.66            Else goto 7.3;
7.67        }while(1)
```

## 4. Conclusions

In this paper we have presented a prevention mechanism against DDoS attack by concurrently performing stateless and stateful signature analysis. The main benefit of combining the two approaches is that only the packets common to both approaches are considered as attack packets and are dropped. Hence legitimate packets remain unaffected by attack. Also only small number of packets need to be inspected for confirmation of attack thus saving the energy while inspection and time of retransmissions. The approach while preventing also helps in tracing the source of attack without applying any specific traceback technique and resources for implementation. Thus attack traffic gradually gets blocked at source and network continues with normal proceedings.

## References

[1]  A. Wood and J. Stankovic. "Denial Of Service In Sensor Networks", IEEE Computer, Vol. 35, Issue: 10, Oct 2002, pp 15-28.
[2]  K. Paul, R.R. Choudhuri, S. Bandyopadhyay, "Survivability Analysis of Ad Hoc Wireless Network Architecture", In proceedings of the IFTP-TC6/European Commission International Workshop on Mobile and Wireless Communication Networks, Springer, Vol. 1818, 2000, pp 31-46
[3]  Andrew P. Snow, Upkar Varshney, Alisha D. Malloy, "Reliability and Survivability of Wireless and Mobile Networks," IEEE Computer, IEEE Computer Society, Vol. 33, Issue: July 2000, pp. 49- 55.
[4]  H. Wang, B. Sheng, C. C. Tan, and Q. Li, "Comparing Symmetric-key and Public-key based Security Schemes in Sensor Networks: A Case Study of User Access Control," In the 28th International Conference on Distributed Computing Systems, Beijing, China, 2008.

[5] A. Ali and N. Fisal, "Security Enhancement For Real-Time Routing Protocol In Wireless Sensor Networks," 5th IFIP International Conference on Wireless and Optical Communications Networks, WOCN '08, 2008.

[6] Deng, R. Han, and S. Mishra, "Defending Against Path-Based Dos Attacks In Wireless Sensor Networks", In ACM SASN '05, November 2005, pp. 89-96.

[7] C. Wan, S. Eisenman, and A. Campbell. "CODA: Congestion Detection And Avoidance In Sensor Networks", In ACM SASN '03, 2003, pp. 266-279.

[8] J. Ioannidis and S. M. Bellovin, "Implementing Pushback: Router-Based Defense Against Ddos Attacks," in Proceedings of Network and Distributed System Security Symposium, NDSS'02, San Diego, CA, 2002, pp. 6-8.

[9] A. Hussain, J. Heodemann, and C. Papadopoulos, "A Framework For Classifying Denial Of Service Attacks," in Proceedings of the 2003 conference on Applications, technologies, architectures, and protocols for computer communications, Karlesruhe, Germany, 2003, pp. 99–110.

[10] A. Kazmanovic and E. W. Knightly, "Low-Rate TCP-Targeted Denial Of Service Attacks," in Proceedings of Symposium, Communication Architecture Protocols, Karlesruhe, Germany, 2003, pp. 345–350.

[11] S. Shyne, A. Hovak, and J. Riolo, "Using Active Networking To Thwart Distributed Denial Of Service Attacks," in Proceedings of IEEE Aerospace Conference, Vol. 3, 2001, pp: 1103-1108

[12] D. Sterne, R. Balupari, W. La Cholter and A. Purtell, "Active Network Based DDoS Defense", in Proceedings of DARPA Active Networks Conference and Exposition, San Francisco, CA, 2002, pp. 193–203.

[13] M. Chiang, Z. Zilic, J. Chenard, K. Radecka, "Architectures of Increased Availability Wireless Sensor Network Nodes", in Proceedings of International Test Conference, IEEE, 2004, pp. 1232-1241.

[14] J. Haggerty, Qi Shi And M. Merabti, "Early Detection And Prevention Of Denial-Of-Service Attacks: A Novel Mechanism With Propagated Traced-Back Attack Blocking", IEEE Journal On Selected Areas In Communications, Vol. 23, No. 10, Oct 2005. pp: 1994-2002

[15] D. Juneja, N. Arora and S. Bansal, "An Ant-Based Routing Algorithm For Detecting Attacks In Wireless Sensor Networks", in International Journal of Computational Intelligence and Research, Research India Publications, Vol. 6, 2010, pp. 311-330.

[16] J.Walters, Z. Liang, W. Shi, and V.Chaudhary, "Wireless Sensor Network Security: A Survey", Chapter 17 Security in Distributed, Grid, and Pervasive Computing. CRC Press, 2006.

[17] J. Bruten, O.Holland and R.Schoonderwoerd, "Ant-Like Agents For Load Balancing In Telecommunications Networks", in Proceedings of 1st International Conference on Autonomous Agent, ACM, 1997, pp. 209-216.

[18] L.Osadciw, R..Muraleedharan and, "Cross Layer Denial of Service Attacks in Wireless Sensor Network Using Swarm Intelligence" In IEEE Long Island Systems, Applications, and Technology Conference (LISAT2007) Farmingdale, New York, May 2007.

[19] L.Osadciw ,R..Muraleedharan, "Decision Making in a Building access system Using Swarm intelligence and Posets", 38th Annual Conference on Information Sciences and Systems, Princeton University, 2004.

[20] L.Osadciw, R..Muraleedharan, "Jamming Attack Detection and Countermeasures In Wireless Sensor Network Using Ant System", SPIE Defence and Security, Orlando, 2006.

[21] C. Meadows, "A Cost-Based Framework For Analysis Of Denial Of Service In Networks", Journal of Computer Security, vol. 9, 2001, pp. 143–164.

[22] J. C. Brustoloni, "Protecting Electronic Commerce From Distributed Denial-of-Service Attacks", in Proceedings of 11$^{th}$ International Conference on World Wide Web, 2002, pp. 553–561.